\documentclass[aps,superscriptaddress,showpacs,floatfix]{revtex4-1}

\usepackage{hyperref}
\usepackage{color}
\usepackage{graphicx} 

\begin{document}

\title{Retarded Correlators in Kinetic Theory: Branch Cuts, Poles and Hydrodynamic Onset Transitions}

\author{Paul Romatschke}
\affiliation{Department of Physics, University of Colorado, Boulder, Colorado 80309, USA}
\affiliation{Center for Theory of Quantum Matter, University of Colorado, Boulder, Colorado 80309, USA}

\begin{abstract}
In this work the collective modes of an effective kinetic theory description based on the Boltzmann equation in relaxation time approximation applicable to gauge theories at weak but finite coupling and low frequencies are studied. Real time retarded two-point correlators of the energy-momentum tensor and the R-charge current are calculated at finite temperature in flat space-times for large N gauge theories. It is found that the real time correlators possess logarithmic branch cuts which in the limit of large coupling disappear and give rise to non-hydrodynamic poles that are reminiscent of quasi-normal modes in black holes. In addition to branch cuts, correlators can have simple hydrodynamic poles, generalizing the concept of hydrodynamic modes to intermediate wavelength. Surprisingly, the hydrodynamic poles cease to exist for some critical value of the wavelength and coupling reminiscent of the properties of onset transitions.
\end{abstract}

\maketitle

\section{Introduction}

Thermal correlators of the energy-momentum tensor and the R-charge current are interesting because they carry information about the momentum and charge diffusion transport properties of the system under consideration. In the case of strongly coupled systems, for which a standard description in terms of weakly coupled quasi-particles is not possible, such information could be important for developing a better understanding of the transport dynamics. Applications to experimentally realizable systems such as high-$T_c$ superconductors, strongly interacting quantum gases and quantum-chromodynamic plasmas provide further motivation to provide a better theoretical understanding of these correlators.

Thermal correlators of large $N$ gauge theories are of particular theoretical interest because this case lends itself to controlled calculations both in the small and large t'Hooft coupling limit via perturbation theory and gauge/gravity duality, respectively (see e.g. Refs.~\cite{Arnold:2003zc,Petreczky:2005nh,Hong:2010at,Moore:2010bu,Teaney:2006nc,Son:2002sd,Starinets:2002br,Kovtun:2005ev,Kovtun:2006pf,CaronHuot:2006te,Myers:2007we,Baier:2007ix,Arnold:2011ja,Kovtun:2012rj}). Furthermore, in this limit, the generically present non-analytic hydrodynamic long time tails are suppressed in the correlators \cite{Kovtun:2003vj,Kovtun:2011np}, simplifying the calculation of thermal correlators both conceptually and technically. Nevertheless, existing results for thermal correlators in the case of large $N$ gauge theories exhibit a highly non-trivial feature for any non-vanishing wave number $k$. Namely, it has been found that while correlators contain branch cuts at vanishing t'Hooft coupling, these cuts are absent at infinite t'Hooft coupling where the correlators contain poles \cite{Hartnoll:2005ju} (at vanishing $k$, one finds poles both at weak and strong coupling). In Ref.~\cite{Hartnoll:2005ju} arguments were given that the branch cut structure should extend to weak, but finite t'Hooft coupling, suggesting some sort of 'phase transition' in the analytic structure of the correlators at a particular, finite value of the t'Hooft coupling. This peculiar finding provides the motivation for revisiting thermal correlators of large $N$ gauge theories in effective kinetic theory in this work. 

Much of the present work is similar to the numerical study in Ref.~\cite{Hong:2010at}. The main difference here is the full analytic treatment made possible by using the relaxation time approximation, which helps elucidate the analytic structure of correlators and the ability to perform limits of weak coupling, strong coupling, and hydrodynamics, respectively.

\section{Collective modes of the Boltzmann equation}

Let us consider the case of a gauge theory such as ${\cal N}=4$ SYM in the limit of a large number of colors $N\gg 1$ at fixed t'Hooft coupling $\lambda=g^2 N$. In the case of weak t'Hooft coupling, a quasiparticle description should be applicable, and at late times the dynamics of the energy-momentum tensor and R-charge, respectively, is then well captured by kinetic theory, cf. Ref.~\cite{Huot:2006ys}. Since the focus of the present work is the analytic structure of correlators rather than their precise numerical evaluation, the relaxation-time approximation to kinetic theory will be employed.  In the relaxation-time approximation, the (exact and complicated) collision integral in kinetic theory is approximated by an effective mean free time $\tau$, or more precisely different relaxation times $\tau_{R,D}$ describing momentum-relaxation through large-angle scattering ($\tau_R$) or charge diffusion ($\tau_D$), respectively. The relaxation-time or Bhatnagar–-Gross–-Krook (BGK, \cite{PhysRev.94.511}) approximation is generally expected to yield results which are qualitatively reliable and even quantitatively accurate.

Ignoring furthermore quantum statistics, which should not affect results qualitatively, one may effectively lump together scalars, vectors and fermions into effective on-shell particle distribution functions $f(t,{\bf x},{\bf p})$ which then define the energy-momentum tensor $T^{\alpha \beta}$ and charge current $J^\alpha$ as
\begin{equation}
\label{eq:defs}
T^{\alpha \beta}(t,{\bf x})=\int \frac{d^3 p}{(2 \pi)^3} \frac{p^\alpha p^\beta}{p^0} f(t,{\bf x},{\bf p})\,,\quad
J^\alpha(t,{\bf x})=q \int \frac{d^3 p}{(2 \pi)^3} \frac{p^\alpha}{p^0} f(t,{\bf x},{\bf p})\,,
\end{equation}
with $q$ the charge of a single quantum. An evolution equation for the distribution function consistent with the conservation of energy, momentum and charge is the Boltzmann equation in the BGK approximation,
\begin{equation}
\label{Boltzmann}
p^\nu \partial_\nu f(t,{\bf x},{\bf p})+F^\alpha \nabla_{\alpha}^{(p)} f(t,{\bf x},{\bf p})=\frac{p^\alpha u_\alpha}{\tau_{R,D}}\left(f-f_{\rm eq}\right)\,,
\end{equation}
where $f_{\rm eq}(t,{\bf x},{\bf p})$ is the (local) equilibrium distribution function that is just a function of the conserved quantities, e.g. local temperature $T(x)$, local chemical potential $\mu(x)$ and local macroscopic velocity $u^\mu(x)$. Note that since the particle distribution function $f$ is take to be on-shell, it does not depend on $p^0$ explicitly, and hence  $\nabla_{0}^{(p)} f=0$. 
$F^{\alpha}$ here is a force originating from an external electromagnetic or gravitational field, and in the case of electromagnetic/gravitational field is given by
$F^\alpha=F^{\alpha\beta}p_\beta$ and $F^{\alpha}=-\Gamma^\alpha_{\beta \gamma}p^\beta p^\gamma$, respectively \cite{Romatschke:2011qp}\footnote{Note that because Eq.~(\ref{Boltzmann}) is written for the on-shell distribution function $f(t,{\bf x},{\bf p})$, the derivative with respect to the energy $\nabla_{0}^{(p)}f$ vanishes since $f$ is independent of $p^0$}. Here $F^{\alpha\beta}$ and $\Gamma^\alpha_{\beta\gamma}$ are the electromagnetic field-strength tensor and the Christoffel symbols, respectively. In the absence of quantum statistics, but in the presence of external fields, the equilibrium distribution function simply is
\begin{equation}
\label{eq:feqfix}
f_{\rm eq}(t, {\bf x},{\bf p})=\exp{\left[\frac{g_{\alpha \beta} p^\alpha u^\beta(t,{\bf x})+\mu(t,{\bf x})}{T}\right]}\,,
\end{equation}
where $\mu(t,{\bf x})$ generally will include contributions from the field $A^\mu$ through the requirement that $f_{\rm eq}$ be a solution to (\ref{Boltzmann}) for a non-vanishing electromagnetic field.

The effective relaxation times $\tau_{R,D}$ in Eq.~(\ref{Boltzmann}) for large angle scattering/charge diffusion have been calculated in the weak-coupling limit from the underlying microscopic collision kernel, finding \cite{Arnold:2003zc}
\begin{equation}
\label{eq:taus}
\tau_{R,D}\propto\frac{1}{\lambda^2 \ln \lambda}\,,\quad
\end{equation}
with proportionality constants that depend on the gauge theory and charge(s) in question. 

\subsection{Retarded Correlators}

The retarded two-point correlation function of a particle current $J^\mu$ and the energy-momentum tensor $T^{\mu \nu}$ are defined in position space as
\begin{eqnarray}
\label{eq:corrdefs}
G^J_{\mu,\nu}(t-t^\prime,{\bf x}-{\bf x}^\prime)&=&-i \theta(t-t^\prime)\langle [J_\mu(t,{\bf x}), J_\nu(t^\prime,{\bf x^\prime})\rangle\,,\nonumber\\
G^T_{\alpha \beta,\gamma \delta}(t-t^\prime,{\bf x}-{\bf x}^\prime)&=&-i \theta(t-t^\prime)\langle [T_{\alpha \beta}(t,{\bf x}), T_{\gamma \delta}(t^\prime,{\bf x^\prime})\rangle\,.
\end{eqnarray}

The canonical approach to calculating thermal correlators is to use linear response theory (see e.g. Ref.~\cite{1963AnPhy..24..419K} for a review of this topic in non-relativistic systems). In the canonical approach, one introduces sources of the conserved quantities (e.g. particle number, energy and momentum) and calculates the response of the hydrodynamic variables in linear response. While calculationally straightforward, the canonical approach has two disadvantages. One disadvantage is that it does not allow direct access to the spatially transverse correlation functions. The other disadvantage is that a naive calculation of the linear response of e.g. the energy-momentum tensor correlators in flat space-time gives wrong results in some cases \cite{PRprep}. Thus, in the following I employ what has been called 'variational approach' to calculate thermal correlation functions which does not suffer from these disadvantages.

In the variational approach, one introduces sources that couple directly to the particle current $J^\mu$ and the energy-momentum tensor $T^{\mu\nu}$, rather than just the conserved quantities, in order to calculate correlation functions. From gauge principle considerations, the source coupling to the particle current has to be the gauge field $A^\mu$, while the source coupling to the energy-momentum tensor has to be the metric field $g^{\mu\nu}$. One-, two-, and higher-point correlation functions of the current and energy-momentum tensor without an electromagnetic field and in flat space may then be defined through expansions of $J^\mu,T^{\mu\nu}$ in the presence of background gauge field and metric perturbations, respectively,
\begin{eqnarray}
\label{eq:Cdefs}
J^\mu(x,A^\nu)&=&G_J^\mu-G_J^{\mu,\nu}\delta A_\nu+\ldots\,,\nonumber\\
T^{\mu\nu}(x,g^{\mu\nu})&=&G_T^{\mu\nu}-\frac{1}{2}G_T^{\mu\nu,\alpha\beta}\delta g_{\alpha \beta}+\ldots\,.
\end{eqnarray}
Using this method, thermal two- and three point correlation functions have been calculated \cite{Baier:2007ix,Moore:2010bu,Arnold:2011ja}. Note that the correlation functions $G^J,G^T$ defined this way will differ from the correlation functions for the energy momentum tensor density $\sqrt{-g}T^{\mu\nu}$ through contact terms (e.g. terms independent of frequency and wave-number in Fourier-space), cf. the discussion in Ref.~\cite{Arnold:2011ja}. Thus, I will ignore any contact terms arising in the following calculations.

\subsection{R-charge diffusion}

Let us first calculate the thermal correlator for R-charge diffusion in a homogeneous static background for which $T(t,{\bf x})=T_0={\rm const}$ and $u^a(t,{\bf x})=(1,0,0,0)$ in the presence of a small external electromagnetic field with gauge potential $\delta A^\mu$. 
Thus, to zeroth order in small perturbations, the evolution equation (\ref{Boltzmann}) requires $f_0=f_{\rm eq,0}=\exp{\left[-\frac{p^0-\mu_0}{T_0}\right]}$. Small perturbations $\delta f$ are required to obey
\begin{equation}
\label{dB}
\left[\partial_t+{\bf v}\cdot {\nabla}\right]\delta f(t,{\bf x},{\bf p})-\frac{f_0}{T_0} {\bf v}\cdot {\bf E}=-\frac{1}{\tau_{D}}\left(\delta f-\delta f_{\rm eq}\right)\,,
\end{equation}
where ${\bf v}\equiv \frac{{\bf p}}{p^0}$ and $\tau_D$ is evaluated at the constant background values of temperature and chemical potential. Here ${\bf E}$ is the electric field defined as ${\bf E}=\nabla A_0-\partial_t {\bf A}$.
  Using the explicit form of $f_{\rm eq}$ in Eq.~(\ref{eq:feqfix}), small perturbations of the equilibrium distribution function can be written as
\begin{equation}
\delta f_{\rm eq}(t,{\bf x},{\bf p})=f_0({\bf p})\frac{\delta \mu(t,{\bf x})}{T_0}\,.
\end{equation}
A Fourier space-transform defined as 
\begin{equation}
\label{eq:FT}
Q(\omega,{\bf k})=\int_{-\infty}^\infty dt \int d^3x e^{i \omega t-i {\bf k}\cdot {\bf x}} Q(t,{\bf x})\,,
\end{equation}
for a quantity $Q(t,{\bf x})$ 
then leads to
\begin{equation}
\label{eq:ftdB}
\delta f(\omega,{\bf k},{\bf p})=\frac{f_0}{T_0}\frac{\delta \mu(\omega,{\bf k})+\tau_D {\bf v}\cdot {\bf E}}{1+\tau_D\left(-i\omega+i{\bf k}\cdot {\bf v}\right)}\,.
\end{equation}
The fluctuating distribution function gives rise to a fluctuating number density of charges $\delta n=\delta J^0/q$ which from Eq.~(\ref{eq:defs}) is given by
\begin{equation}
\label{eq:dneq}
\delta n (\omega,{\bf k})= \int \frac{d^3p}{(2\pi)^3} \delta f(\omega,{\bf k},{\bf p})=\int \frac{dp}{2 \pi^2 T_0} p^2 f_0({\bf p}) \int \frac{d\Omega}{4\pi} \left[
\frac{\delta \mu+\tau_D {\bf v}\cdot {\bf E}}{1+\tau_D\left(-i\omega+i{\bf k}\cdot {\bf v}\right)}
\right]\,.
\end{equation}
This equation may be brought in a more standard form when employing the static susceptibility $\chi\equiv \frac{\partial n}{\partial \mu}$. Since $\delta n=\chi \delta \mu$ thus appears in Eq.~(\ref{eq:dneq}) both on the right-hand-side and the left-hand-side, Eq.~(\ref{eq:dneq}) has the form of a self-consistency equation that must be solved for $\delta n$. Fortunately, in the present case this is trivial since the integral on the rhs in Eq.~(\ref{eq:dneq}) is straightforward. One can recognizing the $dp$ integral to evaluate to the static susceptibility,
\begin{equation}
\chi=\int \frac{dp}{2 \pi^2 T_0} p^2 f_0({\bf p})\,,
\end{equation}
and then solve Eq.~(\ref{eq:dneq}) for $\delta n$ to find
\begin{equation}
\delta n(\omega,{\bf k})=\chi \frac{\int \frac{d\Omega}{4\pi}\frac{\tau_D {\bf v}\cdot {\bf E}}{1+\tau_D\left(-i\omega+i{\bf k}\cdot {\bf v}\right)}}{1-\frac{1}{2 i k \tau_D} \ln \left(\frac{\omega-k+\frac{i}{\tau_D}}{\omega+k+\frac{i}{\tau_D}}\right)}\,.
\end{equation}
For simplicity, let us pick the perturbations to be of the form $\delta n(t,x_3)$ in the following, e.g. select the momentum vector ${\bf k}$ to point along the $x_3$ direction. The retarded thermal correlator for the particle current is then found from calculating $G^{0,0}_J(\omega,k)=\frac{\delta n}{\delta A^0}$, $G^{1,1}_J(\omega,k)=\frac{\delta J^1/q}{\delta A^1}$,  etc. One finds
\begin{eqnarray}
\label{eq:Gnnres}
G^{0,0}_J(\omega,k)&=&-\chi \frac{1+(i \omega -\frac{1}{\tau_D})\frac{1}{2 i k} \ln \left(\frac{\omega-k+\frac{i}{\tau_D}}{\omega+k+\frac{i}{\tau_D}}\right)}{1-\frac{1}{2 i k \tau_D} \ln \left(\frac{\omega-k+\frac{i}{\tau_D}}{\omega+k+\frac{i}{\tau_D}}\right)}\,,\\
G^{0,3}_J(\omega,k)&=&-\frac{\omega \chi}{k}\frac{1+(i \omega -\frac{1}{\tau_D})\frac{1}{2 i k} \ln \left(\frac{\omega-k+\frac{i}{\tau_D}}{\omega+k+\frac{i}{\tau_D}}\right)}{1-\frac{1}{2 i k \tau_D} \ln \left(\frac{\omega-k+\frac{i}{\tau_D}}{\omega+k+\frac{i}{\tau_D}}\right)}\,,\nonumber\\
G^{1,1}_J(\omega,k)&=&\frac{\chi i \omega \tau_D}{4}\left[
\frac{2 (1-i \omega \tau_D)}{k^2 \tau_D^2}+\frac{(1-i \omega \tau_D)^2+k^2 \tau_D^2}{(i \tau_D k)^3} \ln \left(\frac{\omega-k+\frac{i}{\tau_D}}{\omega+k+\frac{i}{\tau_D}}\right)
\right]\,,\nonumber
\end{eqnarray}
etc. The above results fulfill the Ward identities following from $\partial_\mu J^\mu=0$, e.g.
$$
\omega G_J^{0,0}(\omega,k)=k G_J^{0,3}(\omega,k)\,,\quad
\omega^2 G_J^{0,0}(\omega,k)=k^2 G_J^{3,3}(\omega,k)\,.
$$

\subsubsection{Hydrodynamic Limit}

The limit of hydrodynamics corresponds to considering small frequencies and small wave-numbers. The correlation functions $G_J^{\mu,\nu}$ in the hydrodynamic limit may be obtained in a manner fully equivalent to those described above. Including terms up to second-order in gradients, one finds  for instance \cite{Hong:2010at,PRprep,ANDP:ANDP19003060312}
\begin{equation}
\label{eq:grjjhydro}
G_{J,{\rm hydro}}^{0,0}(\omega,k)=\frac{k^2(\chi D)}{i\omega(1-i \tau_\Delta \omega)- D k^2}\,,
\end{equation}
where $D$ is the usual particle number diffusion constant, and $\tau_\Delta$ is the 'diffusion relaxation time' which controls the time-scale on which dissipative stresses relax to Landau-Lifshitz type diffusion dynamics (not to equilibrium). Eq.~(\ref{eq:grjjhydro}) generalizes the known hydrodynamic result \cite{Kovtun:2012rj} to second-order hydrodynamics where diffusion is described by a causal process (cf. Refs.~\cite{Hiscock:1985zz,Romatschke:2009im}). It should be stressed that while Eq.~(\ref{eq:grjjhydro}) matches the so-called Drude model result \cite{ANDP:ANDP19003060312}, the derivation in terms of hydrodynamics does at no point assume the presence of weakly coupled quasiparticles, and hence is applicable even in the strong-coupled regime where quasiparticle degrees of freedom are absent.

Since hydrodynamics is the universal long-wavelength, low-frequency limit of any system with the same symmetries, Eq.~(\ref{eq:grjjhydro}) can be used to derive Kubo relations for the value of the hydrodynamic transport coefficients $D,\tau_\Delta$, e.g.
\begin{eqnarray}
\label{eq:dkubo}
\lim_{\omega\rightarrow 0}\lim_{k\rightarrow 0}\frac{\omega}{k^2}{\rm Im}\,G^{0,0}_{J}(\omega,{\bf k})&=& - \chi D\,,\nonumber\\
\lim_{\omega\rightarrow 0}\lim_{k\rightarrow 0}\frac{1}{k^2}{\rm Re}\,G^{0,0}_{J}(\omega,{\bf k})&=&  \chi D\tau_\Delta\,.
\end{eqnarray}

Applying these Kubo formulas to the result found in kinetic theory (\ref{eq:Gnnres}), one finds the diffusion constant and second-order transport coefficient $\tau_\Delta$ in terms of the diffusion time scale $\tau_D$:
\begin{equation}
\label{eq:Dkubo}
D=\frac{\tau_D}{3}\,,\quad \tau_\Delta=\tau_D\,.
\end{equation}
Thus the kinetic theory timescale $\tau_D$ plays a double role through controlling both the transport coefficient $D$ as well as the hydrodynamic relaxation time $\tau_\Delta$ (this is an artifact stemming from the single-relaxation time ansatz in Eq.~(\ref{dB}), which does not affect the conclusions in this work).
In the hydrodynamic limit, one may be tempted to cast Eq.~(\ref{eq:Gnnres}) into the form of Eq.~(\ref{eq:grjjhydro}). To study the pole structure of the correlator, one can expand the inverse of $G_{J}^{0,0}(\omega,{\bf k})$ in a power series of $k$, obtaining
\begin{equation}
\label{eq:gnnhydro}
G^{0,0}_J(\omega,{\bf k})=\frac{\chi k^2  (1- i \tau_D \omega)\tau_D/3}{i \omega(1-i \tau_D \omega)^2-(1-\frac{9}{5} i \tau_D \omega) k^2 \tau_D/3+{\cal O}({\bf k}^3)}\,,
\end{equation}
which is similar to, but does not quite match, Eq.~(\ref{eq:grjjhydro}). The differences between a low-momentum expansion of Eq.~(\ref{eq:Gnnres}) and Eq.~(\ref{eq:grjjhydro}) are beyond the accuracy of second-order hydrodynamics, so they do not represent an inconsistency. One finds that Eq.~(\ref{eq:gnnhydro}) has three poles in the small $k$ limit, located at 
\begin{equation}
\label{eq:simpRpoles}
\omega_0^{(D)}=-i \frac{\tau_D}{3} k^2+{\cal O}(k^3)\,,\quad
\omega_{1,2}^{(D)}=-\frac{i}{\tau_D}\pm  \frac{2k}{\sqrt{15}} +{\cal O}(k^2)\,.
\end{equation}
The pole located at $\omega=\omega_0^{(D)}$ can be recognized as the usual hydrodynamic diffusion pole, while the poles at $\omega=\omega_{1,2}^{(D)}$ are ``non-hydrodynamic'' poles, as they do not obey $\lim_{k\rightarrow 0}\omega_{1,2}^{(D)}(k)\rightarrow 0$. In fact, $\omega_{1,2}^{(D)}$ somewhat resemble the dispersion relation found for quasi-normal modes in black holes. (See Ref.~\cite{Brewer:2015ipa} for a discussion on experimental hints for non-hydrodynamic modes in cold quantum gases).

\subsubsection{Weak Coupling Limit}

In the weak coupling limit, one has $\tau_D T \gg 1$. It turns out that the correlators $G_J$ depend on combinations such as $\tau_D k=\tau_D T \frac{k}{T}$ rather than $\tau_D T$ alone. Thus, the weak coupling limit taken here corresponds to the case where the ratios of wavenumber and frequency to temperature, $\frac{k}{T},\frac{\omega}{T}$ are held constant while $\tau_D T\rightarrow \infty$. Thus, formally, in this case the weak coupling limit coincides with both the short wavelength limit (large wavenumber and frequency limit) as well as the zero temperature limit where $\frac{\omega}{T}\gg 1,\frac{k}{T}\gg1$. As a consequence, for the Boltzmann equation, the weak coupling, large frequency/wavenumber and zero temperature limits lead to identical results, which has been used before (see e.g. Ref.~\cite{Hong:2010at}). Some aspects of the similarity of these limits can also be observed in quantum field theories, see e.g. calculations of the large frequency and zero temperature limits in gauge/gravity calculation \cite{Son:2002sd,Teaney:2006nc} which qualitatively and in some cases quantitatively match those from free field theory. In the case at hand, one finds for the weak coupling limit $\tau_D T\gg 1$
\begin{equation}
\label{eq:gjjweak}
\lim_{\tau_D\rightarrow \infty}G_J^{0,0}(\omega,{\bf k})=-\chi\left[1+\frac{\omega}{2k}\ln \left(\frac{\omega-k}{\omega+k}\right)+\frac{i \omega}{\tau_D (\omega^2-k^2)}-\frac{ i \omega}{4 k^2\tau_D} \ln^2 \left(\frac{\omega-k}{\omega+k}\right)+{\cal O}\left(\frac{1}{\tau_D^2}\right)\right]\,.
\end{equation}
In the limit of a free theory ($\tau_D\rightarrow \infty$) one has ${\rm Im}G_J^{0,0}(\omega,k)=-\pi \chi \frac{\omega}{2 k}\theta(k-\omega)\theta(k+\omega)$, where $\theta(x)$ is the step function. From this result it is easy to show that $\lim_{k\rightarrow 0}\frac{{\rm Im}G_J^{0,0}(\omega,k)}{\omega}$ gives a representation of the Dirac $\delta$ function and one obtains ${\rm Im}G_J^{0,0}(\omega,k=0)=-\pi \chi \omega \delta(\omega)$, matching the result derived in the appendix of Ref.~\cite{Petreczky:2005nh}. 

Eq.~(\ref{eq:gjjweak}) implies the presence of a logarithmic branch cut originating from singularities at $\omega=\pm k$ in the weak-coupling limit of $G_J^{0,0}$. This is unlike the pole structure found in the hydrodynamic limit of the same correlator discussed above.

One should note that Eq.~(\ref{eq:gjjweak}) does not contain a term proportional to $\omega^4$ that can be derived from a free (classical) field theory calculation (see e.g. \cite{Teaney:2006nc}). The reason for this is that when a kinetic equations are derived from quantum field theories, a gradient-expansion step is included when performing the Wigner-transforms (see e.g. \cite{Blaizot:2001nr}). This implies that the kinetic theory result does not accurately capture terms of order $\omega^4$.

\subsubsection{Strong Coupling Limit}

The dimensionless product of relaxation time and temperature $\tau_D T$ becomes smaller as the coupling is increased from zero. Naively continuing this process with a relaxation time given by Eq.~(\ref{eq:taus}) would imply $\tau_D T \rightarrow 0$ in the limit of large t'Hooft coupling $\lambda\rightarrow \infty$. It should be noted that in actual ${\cal N}=4$ SYM theory the strong coupling limit of $\tau_D T$ remains small but finite, which is the reason for calling the limit $\tau_D T \rightarrow 0$ a 'naive' strong coupling limit (see section \ref{sec:three} for more discussion on this topic). Also, for reasons discussed above, the dependence of the correlators $G_J$ on $\tau_D$ should be understood in the sense of keeping the ratios $\frac{k}{T},\frac{\omega}{T}$ fixed while sending $\tau_D T \rightarrow 0$. In order to obtain an approximating function to $G_J^{0,0}(\omega,{\bf k})$ in this limit, it is useful to employ a Pad\'e approximant
\begin{equation}
\label{eq:Padedef}
R_{m}(\omega,{\bf k})\equiv \frac{\sum_{i=0}^m u_i(\omega,{\bf k})\tau_D^i}{\sum_{i=0}^m d_i(\omega,{\bf k})\tau_D^i}\,,
\end{equation}
where for simplicity only symmetric Pad\'e approximants are considered. The need for a Pad\'e approximation originates from the known behavior of $G_{J}$ as a ratio of two functions, rather than just a simple power series expansion. 
Requiring $R_{m}(\omega,{\bf k})$ to match $G_J^{0,0}(\omega,{\bf k})$ order by order in a power series expansion in $\tau_D$ leads to consistency equations determining the coefficients $u_i,d_i$. Solving these equations for $m=2$ leads to
\begin{equation}
\label{eq:Gnnstrong}
\lim_{\tau_D\rightarrow 0}G_J^{0,0}(\omega,{\bf k})\simeq R_{2}(\omega,{\bf k})=
\frac{\chi k^2 \tau_D/3 (1- i \tau_D \omega)}{i \omega(1-i \tau_D \omega)^2-k^2 \tau_D/3(1-\frac{9}{5} i \tau_D \omega)}\,,
\end{equation}
which precisely matches the result in the hydrodynamic limit, Eq.~(\ref{eq:gnnhydro}). Therefore, the analytic structure of the correlator in the strong coupling limit is such that it contains only poles, but no branch-cuts. The formal correspondence of hydrodynamic (long wavelength) limit and strong coupling limit is again a direct consequence of the fact that $G_J$ for the Boltzmann equation only depends on $\tau_D$ through combinations such as $\tau_D \omega,\tau_D k$.

\subsection{Momentum Transport}

Let us now calculate the thermal correlator for the energy-momentum tensor in
the background of an arbitrary (but small) metric perturbation $\delta g_{ab}$. The metric perturbation will source perturbations in the temperature and fluid velocity so that $T(t,{\bf x})=T_0+\delta T(t,{\bf x})$, and $u^a(t,{\bf x})=(1,0,0,0)+\delta u^a(t,{\bf x})$, while not coupling to the particle number so $\mu(t,{\bf x})=0$. Solving Eq.~(\ref{Boltzmann}) to leading order for a small perturbation implies $f_0=f_{\rm eq,0}=\exp{\left[-\frac{p^0}{T_0}\right]}$ and hence $\delta f$ has to fulfill the equation
\begin{equation}
\label{dBT}
\left[\partial_t+{\bf v}\cdot {\nabla}\right]\delta f(t,{\bf x},{\bf p})-\Gamma^{\alpha}_{\beta \gamma} \frac{p^\beta p^\gamma}{p^0}\nabla_\alpha^{(p)}f_0=-\frac{1}{\tau_{R}}\left(\delta f-\delta f_{\rm eq}\right)\,.
\end{equation}
Fluctuations in the equilibrium distribution function can be expanded to yield
\begin{equation}
\label{eq:feqhydro}
\delta f_{\rm eq}(t,{\bf x},{\bf p})=f_0({\bf p})\frac{p^0}{T_0}\left({\bf v}\cdot \delta {\bf u}(t,{\bf x})+\frac{\delta T(t,{\bf x})}{T_0}\right)\,,
\end{equation}
where it should be noted that the $\delta g_{00}$ contribution to $u_0=-1+\frac{\delta g_{00}}{2}$ has canceled against the on-shell condition $p^0=|{\bf p}|\left(1+\frac{\delta g_{00}}{2}\right)$.
The fluctuating distribution function gives rise to a fluctuating energy momentum tensor $\delta T^{\mu\nu}$ which from Eq.~(\ref{eq:defs}) is given in Fourier space as\footnote{Note that $\Gamma^i_{\alpha \beta}p^\alpha p^\beta p_i=\Gamma^{0}_{\alpha \beta}p^\alpha p^\beta p_0$ because $\Gamma_{\alpha \beta\gamma}p^\alpha p^\beta p^\gamma=p^\mu \partial_\mu \left(p^2\right)=0$ for on-shell particles.}
\begin{equation}
\label{eq:dtab}
\delta T^{\mu\nu}(\omega,{\bf k})=\int \frac{d p}{2 \pi^2 T_0} p^4 f_0({\bf p})\int \frac{d\Omega}{4 \pi} v^\mu v^\nu \frac{{\bf v}\cdot \delta{\bf u}+\frac{\delta T}{T_0}-\tau_R \Gamma^0_{\alpha \beta}v^\alpha v^\beta}{1+\tau_R\left(-i\omega+i{\bf k}\cdot {\bf v}\right)}\,,
\end{equation}
where $v^\alpha\equiv(1,{\bf v})$. For massless particles, it is easy to recognize $\int \frac{d p}{2 \pi^2 T_0} p^4 f_0({\bf p})=3 (\epsilon_0+P_0)$. One finds that Eq.~(\ref{eq:dtab}) leads to a set of self-consistency equations for temperature and velocity perturbations, defined through
$\delta \epsilon \equiv \delta T^{00}$ and $\delta u^i\equiv\frac{1}{(\epsilon_0+P_0)}\delta T^{0i}$. These can be solved in a straightforward manner when restricting to metric perturbations of the type $\delta g_{ab}(t,x_3)$. One then obtains the retarded correlators in the spin 2 (metric perturbations $\delta g_{12}$), spin 1 (shear, $\delta g_{01},\delta g_{02},\delta g_{13},\delta g_{23}$) and spin 0 (sound, $\delta g_{00},\delta g_{03},\delta g_{11},\delta g_{22},\delta g_{33}$) channels, respectively. One finds
\begin{eqnarray}
\label{eq:allhydrores}
G_T^{12,12}(\omega,k)&=&\frac{3 i \omega \tau_R (\epsilon_0+P_0)}{16}
\left[\frac{10}{3}\frac{1-i \tau_R \omega}{k^2 \tau_R^2}+\frac{2 (1-i \tau_R \omega)^3}{k^4 \tau_R^4}+i\frac{\left((1-i \tau_R \omega)^2+k^2 \tau_R^2\right)^2}{k^5 \tau_R^5} L \right]\,,\nonumber\\
G_T^{01,01}(\omega,k)&=&-(\epsilon_0+P_0)\frac{2 k \tau_R\left(2 k^2 \tau_R^2+3 (1-i \tau_R \omega)^2\right)+3 i (1-i\tau_R \omega)\left(k^2 \tau_R^2+(1-i \tau_R \omega)^2\right) L}{2 k \tau_R (3+2 k^2 \tau_R^2-3 i \tau_R \omega)+3 i \left(k^2 \tau_R^2+(1-i \tau_R \omega)^2\right) L}\,,\nonumber\\
G_T^{13,13}(\omega,k)&=&\frac{\omega^2}{k^2}G_T^{01,01}(\omega,k)\,,\nonumber\\
G_T^{00,00}(\omega,k)&=&-3(\epsilon_0+P_0)\left[1+k^2 \tau_R \frac{2 k \tau_R+i(1-i \tau_R\omega) L}{2 k \tau_R(k^2 \tau_R +3 i \omega)+i \left(k^2 \tau_R+3\omega(i+\tau_R \omega)\right) L}\right]\,,\nonumber\\
G_T^{03,00}(\omega,k)&=&-3(\epsilon_0+P_0)\frac{\omega}{k}\left[k^2 \tau_R \frac{2 k \tau_R+i(1-i \tau_R\omega) L}{2 k \tau_R(k^2 \tau_R +3 i \omega)+i \left(k^2 \tau_R+3\omega(i+\tau_R \omega)\right)  L}\right]\,,\nonumber\\
G_T^{03,03}(\omega,k)&=&-3(\epsilon_0+P_0)\left[\frac{1}{3}+\omega^2 \tau_R \frac{2 k \tau_R+i(1-i \tau_R\omega) L}{2 k \tau_R(k^2 \tau_R +3 i \omega)+i \left(k^2 \tau_R+3\omega(i+\tau_R \omega)\right) L}\right]\,.\nonumber\\
\end{eqnarray}
where the shorthand notation
$$
L=\ln \left(\frac{\omega-k+\frac{i}{\tau_R}}{\omega+k+\frac{i}{\tau_R}}\right)\,,
$$
was used. As was the case for the correlators in the R-charge diffusion channel, one can verify that the above correlators fulfill the Ward identities resulting from $\partial_\mu T^{\mu\nu}=0$, e.g.
$$
\omega G_T^{00,00}(\omega,k)=k G_T^{03,00}(\omega,k)\,,\quad
\omega^2 G_T^{00,00}(\omega,k)=k^2 G_T^{03,03}(\omega,k)\,,
$$
up to contact terms.

\subsubsection{Hydrodynamic Limit}

For second-order hydrodynamics \cite{Baier:2007ix,Bhattacharyya:2008jc,Romatschke:2009kr}, one can calculate the momentum-transport correlation functions in a manner similar to that above. Neglecting contact terms, one finds for instance \cite{PRprep} 
\begin{eqnarray}
\label{eq:GThydro}
G_{T,{\rm hydro}}^{00,00}(\omega,k)&=&\frac{k^2\left[(\epsilon_0+P_0)-k^2\frac{2 \kappa}{3(1-i \tau_\pi \omega)}
\right]}{\omega^2-c_s^2 k^2+ i \omega k^2 \gamma_s(\omega)}\,,\\
G_{T,{\rm hydro}}^{01,01}(\omega,k)&=&\frac{k^2\left(\eta -i \omega \frac{\kappa}{2}\right)}{i \omega(1-i \tau_\pi \omega)-\gamma_\eta k^2}\nonumber\\
G_{T,{\rm hydro}}^{12,12}(\omega,k)&=&-\frac{i \eta \omega+\frac{\kappa}{2}(k^2+\omega^2)}{1-i\tau_\pi \omega}\,,\nonumber\\
\gamma_s(\omega)&=&\frac{4 \eta}{3 (\epsilon_0+P_0) (1-i \tau_\pi \omega)}
\,,\nonumber
\end{eqnarray}
where $\gamma_\eta\equiv \frac{\eta}{\epsilon_0+P_0}$ and $c_s=\sqrt{\frac{d\epsilon_0}{dP_0}}$ is the local speed of sound. Furthermore, $\eta$ is the shear viscosity coefficient and $\tau_\pi$ is the hydrodynamic shear relaxation times that control the time-scales on which dissipative stresses relax to usual Navier-Stokes hydrodynamics (not to equilibrium). $\kappa$ is a second-order transport coefficients which can to be obtained for theories such as QCD and ${\cal N}=4$ SYM, cf. Ref.~\cite{Baier:2007ix,Romatschke:2009ng,Philipsen:2013nea}. Eqns.~(\ref{eq:GThydro}) generalize the well-known results in Navier-Stokes hydrodynamics \cite{Kovtun:2012rj} to second-order conformal hydrodynamics where all dynamics is described by a causal process (cf. Refs.~\cite{Hiscock:1985zz,Romatschke:2009im}). Re-expanding Eqns.~(\ref{eq:GThydro}) in powers of $\omega,k$ one recovers the results found in Ref.~\cite{Arnold:2011ja}.

Momentum transport correlators must match to the universal hydrodynamic forms (\ref{eq:GThydro}) in the limit of small wave-number and small frequency, leading to 
Kubo relations such as \cite{Baier:2007ix}
\begin{eqnarray}
\label{eq:kubohydro}
G_T^{12,12}(\omega,k)&=&-i \eta \omega +\omega^2 \left(\eta \tau_\pi-\frac{\kappa}{2}\right)-\frac{\kappa}{2}k^2+{\cal O}(\omega^3,k^3)\,,
%
%
\end{eqnarray}

Applying the Kubo formula (\ref{eq:kubohydro}) to the momentum transport correlators $G_T^{12,12}$ from the Boltzmann equation (\ref{eq:allhydrores}), one finds
\begin{equation}
\frac{\eta}{s}=\frac{\tau_R T}{5}\,, \quad \tau_\pi=\tau_R\,,\quad \kappa=0\,.
\end{equation}
Thus, as was the case for the R-charge diffusion discussed above, $\tau_R$ takes on a double role as controlling both the hydrodynamic viscosity and relaxation time coefficients.
As before, one may be tempted to recast (\ref{eq:allhydrores}) in forms similar to Eq.~(\ref{eq:GThydro}) in the hydrodynamic limit. For the sound channel correlators, dropping contact terms, an expansion for $\omega\propto k\ll 1$ for the inverse of $G_T^{00,00}$ leads to
\begin{equation}
\label{eq:gt0000hydroa}
G_T^{00,00}(\omega,k)\simeq (\epsilon_0+P_0)\frac{k^2}{\left[\omega^2-k^2\left(\frac{1}{3}-\frac{4 i \tau_R \omega}{15}(1+i \tau_R \omega)\right)+{\cal O}(\omega^5,\omega^4 k,\ldots)\right]}\,,
\end{equation}
which --- to the order of the expansion --- matches the hydrodynamic result (\ref{eq:GThydro}) with $\gamma_s(\omega)=\frac{4 \tau_R}{15 (1-i \tau_R \omega)}$. One finds that Eq.~(\ref{eq:gt0000hydroa}) has two poles, which in the small $k$ limit are located at 
\begin{equation}
\label{eq:soundhydropoles}
\omega^{(0)}_{\pm}=\pm c_s k-\frac{2 i \tau_R k^2}{15}\,.
\end{equation} 
These poles can be recognized as the usual hydrodynamic poles.

For the shear channel, a similar expansion with $k\ll 1$ for the inverse of $G_T^{01,01}$ leads to
\begin{equation}
\label{eq:gt0101hydroa}
G_T^{01,01}(\omega,k)\simeq (\epsilon_0+P_0)\frac{k^2 \tau_R (1-i \tau_R \omega)}{5 i \omega (1-i \tau_R \omega)^2-k^2 \tau_R\left(1-\frac{15}{7} i \tau_R \omega\right)+{\cal O}(k^4)}\,,
\end{equation}
which matches the hydrodynamic result (\ref{eq:GThydro}) with $\gamma_\eta=\frac{\eta}{\epsilon_0+P_0}=\frac{\tau_R}{5}$. One finds that Eq.~(\ref{eq:gt0101hydroa}) implies three poles which in the small $k$ limit are located at 
\begin{equation}
\label{eq:spin1poles}
\omega_{0}^{(1)}=-\frac{i \tau_R k^2}{5}\,,\quad \omega_{1,2}^{(1)}=-\frac{i}{\tau_R}\pm  \tau_R k \sqrt{\frac{8}{35}}\,.
\end{equation}
Again, $\omega_0^{(1)}$ can be recognized to be the usual hydrodynamic diffusion pole, and $\omega^{(1)}_{1,2}$ correspond to two non-hydrodynamic poles.

Finally, an expansion $k\ll 1$ for the inverse of $G_T^{12,12}$ leads to
\begin{equation}
G_T^{12,12}(\omega,k)=3(\epsilon_0+P_0)\frac{\tau_R \omega (1-i \tau_R \omega)}{15\left(i (1-\tau_R \omega)^2+\frac{i k^2 \tau_R^2}{7}\right)+{\cal O}(k^4)}\,,
\end{equation}
which has two non-hydrodynamic poles which in the small $k$ limit are located at 
$$
\omega_{1,2}^{(2)}=-\frac{i}{\tau_R}\pm k \sqrt{\frac{1}{7}}\,.
$$
There are no hydrodynamic poles in the spin 2 channel, matching previous results.

\subsubsection{Weak Coupling Limit}

In the weak coupling limit $\tau_R T \gg 1$ with $\frac{\omega}{T},\frac{k}{T}$ fixed one obtains
\begin{eqnarray}
G_T^{00,00}(\omega,k)&\rightarrow&-6 (\epsilon_0+P_0)\left[1+\frac{\omega}{4 k}\ln \left(\frac{\omega-k}{\omega+k}\right)\right]+{\cal O}(\tau_R)\,,\nonumber\\
G_T^{01,01}(\omega,k)&\rightarrow& \frac{ (\epsilon_0+P_0)}{4 k^3}\left[
-4 k^3+6 k \omega^2+3 \omega(\omega^2-k^2)\ln \left(\frac{\omega-k}{\omega+k}\right)\right]+{\cal O}(\tau_R)\,,\\
G_T^{12,12}(\omega,k)&\rightarrow& \frac{(\epsilon_0+P_0)}{16 k^5}\left[10 k^3 \omega-6 k \omega^3-3 (k^2-\omega^2)^2 \ln \left(\frac{\omega-k}{\omega+k}\right)\right]+{\cal O}(\tau_R)\,,\nonumber
\end{eqnarray}
exhibiting again the now familiar branch-cut analytic structure for the correlators at weak coupling.

\subsubsection{Strong Coupling Limit}

Using again a Pad\'e approximant (\ref{eq:Padedef}) to recover the strong coupling limit of the momentum-transport correlators, dropping contact terms, one finds
\begin{eqnarray}
G_T^{00,00}(\omega,k)&\rightarrow&(\epsilon_0+P_0)\frac{k^2(1-i \tau_R \omega)}{\omega^2(1-i \tau_R \omega)-\frac{k^2}{3}(1-\frac{9}{5} i\tau_R \omega)}\,,\nonumber\\
G_T^{01,01}(\omega,k)&\rightarrow&(\epsilon_0+P_0)\frac{k^2 \tau_R (1-i \tau_R \omega)}{5 i \omega (1-i \tau_R \omega)^2-k^2 \tau_R\left(1-\frac{15}{7} i \tau_R \omega\right)}\,,\\
G_T^{12,12}(\omega,k)&\rightarrow&3 (\epsilon_0+P_0)\frac{\tau_R \omega (1-i \tau_R \omega)}{15\left(i (1-\tau_R \omega)^2+\frac{i k^2 \tau_R^2}{7}\right)}\,,\nonumber
\end{eqnarray}
matching hydrodynamic results (upon re-expanding $\omega\propto k\ll 1$ for the sound channel). Thus, in the strong coupling limit, the momentum-transport correlators only possess poles, but no branch cuts.

\section{Branch cuts, Poles and Hydrodynamic Onset Transitions}
\label{sec:three}

In the previous section, I have derived results for thermal correlators for R-charge diffusion and momentum transport from kinetic theory. The results found for these correlators all possess the following analytic structure for non-zero wave-number $k$:
\begin{itemize}
\item
At weak coupling, correlators exhibit a logarithmic branch cut originating from the singularities located at $\omega=-k$ and $\omega=k$. This is consistent with weak coupling quantum field theory results \cite{Kraemmer:2003gd} and plasma physics where this branch cut is responsible for Landau-damping.
\item
Pushing the results to the strong coupling regime through taking the naive limit $\tau T\rightarrow 0$, correlators exhibit only poles. Some of these poles correspond to the familiar hydrodynamic poles, but in general there are other, non-hydrodynamic poles present that are reminiscent of (but not equivalent to) quasi-normal modes in black holes (cf. Ref.~\cite{Brewer:2015ipa}).
\item
The strong coupling expansion of the kinetic theory result exactly matches the result expected from hydrodynamics for the correlators. This is a generalization of the well-known fact that the dynamics of the one-point function in kinetic theory turns into hydrodynamics in the strong coupling limit (cf. Ref.~\cite{Romatschke:2011qp}).
\end{itemize}

Correlators obtained in kinetic theory seem to exhibit the same qualitative changes in analytic structure observed in exact quantum-field theory calculations in ${\cal N}=4$ SYM: branch cuts at weak coupling to poles at strong coupling. 
For this reason, it is interesting to study the changes in the analytic structure of thermal correlators obtained in kinetic theory, even if the quantitative applicability of kinetic theory remains firmly limited to the weak-coupling region. Because the results for R-charge correlators and momentum-transport correlators were qualitatively similar, I restrict myself to mostly discussing on the R-charge correlation function for simplicity.

\subsection{R-charge diffusion}

\begin{figure}[t]
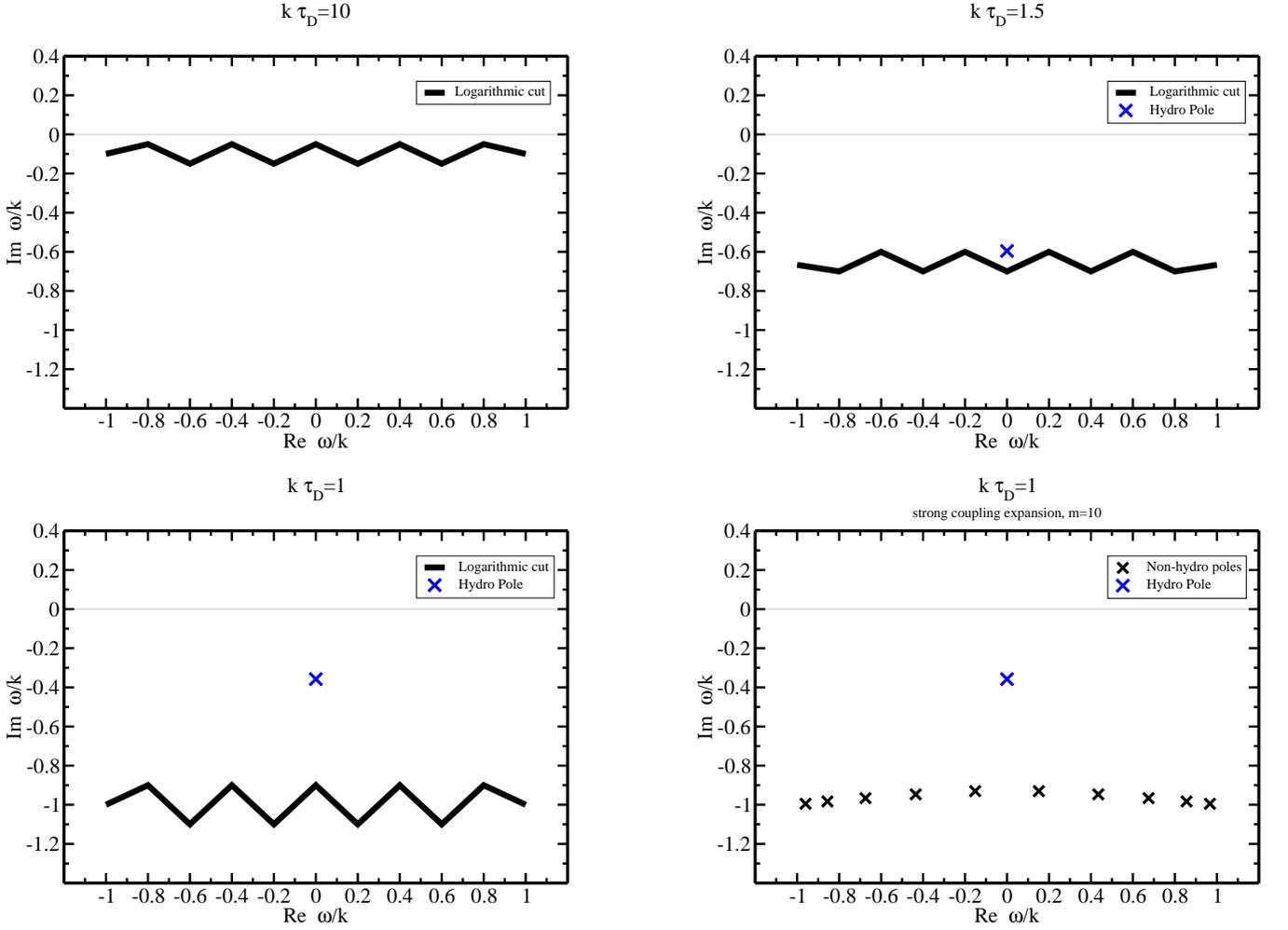

\centering
\includegraphics[width=0.45\textwidth]{fig1a}\hfill
\includegraphics[width=0.45\textwidth]{fig1b}\\
\quad\\
\includegraphics[width=0.45\textwidth]{fig1c}\hfill
\includegraphics[width=0.45\textwidth]{fig1d}
\caption{Analytic structure of the R-current correlator at weak coupling (upper left), intermediate coupling (upper right and lower left) and in the strong-coupling expansion (lower right). Note that the hydrodynamic pole ceases to exist at weak coupling. See text for details.\label{fig:one}}
\end{figure}

Considering the result for the R-charge correlator $G_J^{0,0}(\omega,k)$ in Eq.~(\ref{eq:Gnnres}) at non-vanishing wave-number $k$, one finds that it has a logarithmic branch-cut in the lower-half complex $\omega$ plane, which can be taken to lie in between $\omega=-\frac{i}{\tau_D}-k$ and $\omega=-\frac{i}{\tau_D}+k$. In addition to this branch-cut, $G_J^{0,0}(\omega,k)$ for $k \tau_D <\frac{\pi}{2}$ also possesses a pole in the lower-half complex plane located at 
\begin{equation}
\label{eq:hydropole}
\omega(k)=-\frac{i}{\tau_D}+\frac{i k}{\tan(k \tau_D)}\,.
\end{equation}
The analytic structure of $G_J^{0,0}(\omega,k)$ is sketched in Fig.~\ref{fig:one}. In the weak coupling limit $\tau_D\rightarrow \infty$ the pole in $G_J^{0,0}(\omega,k)$ ceases to exist as it merges with the cut at $\tau_D=\frac{\pi}{2 k}$. The branch cut moves closer to the real axis and becomes the branch cut familiar from Landau-damping, stretching from $\omega=-k$ to $\omega=k$ (see Fig.~\ref{fig:one})

\begin{figure}[t]
\centering
\includegraphics[width=0.49\textwidth]{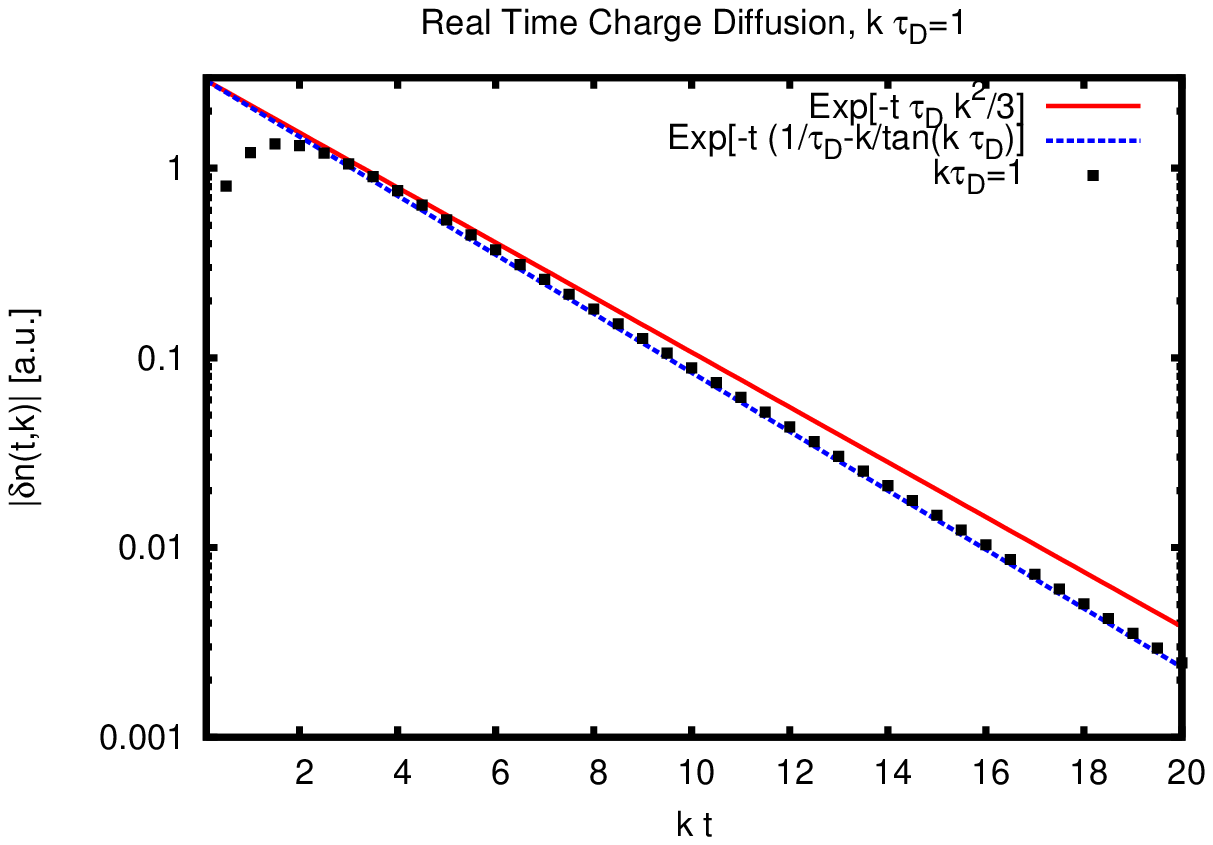}\hfill
\includegraphics[width=0.49\textwidth]{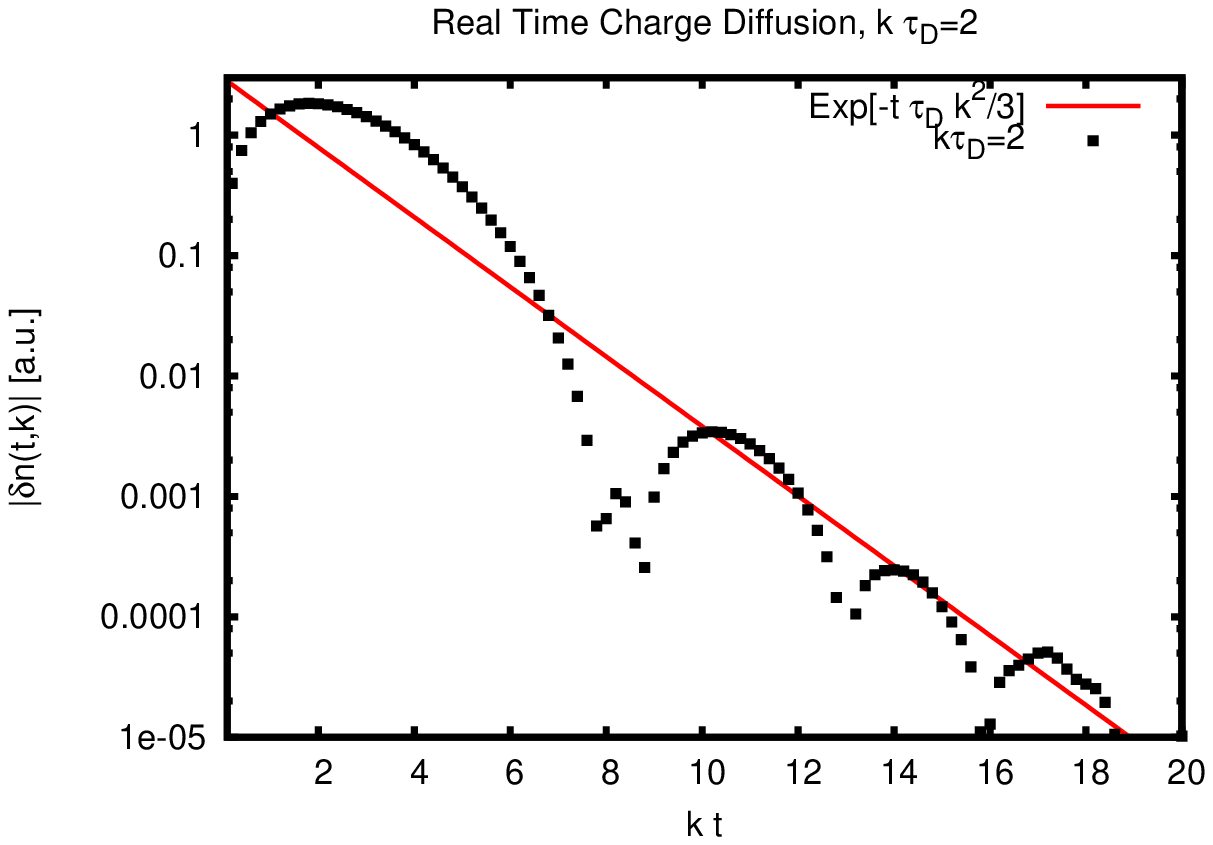}
\caption{Real time evolution of charge at fixed $k$ in the H phase ($k \tau_D=1$, left)  and the G phase ($k \tau_D=2$, left). In both panels, the standard diffusion time evolution $\delta n(t,k)\propto e^{-t \tau_D k^2/3}$ is shown. In the left panel, the actual evolution expected from the hydrodynamic pole (\ref{eq:hydropole}) is shown, while this pole is absent in the G phase. Note the qualitative difference of the time evolution of $\delta n(t,k)$ when crossing the onset transition boundary located at $k \tau_D=\frac{\pi}{2}$.
\label{fig:two}}
\end{figure}

In the naive strong coupling limit $\tau_D T\rightarrow 0$, the pole in $G_J^{0,0}(\omega,k)$ is located at $\omega_0^{(D)}=-\frac{i \tau_D k^2}{3}$ and thus becomes recognizable as the hydrodynamic pole in Eq.~(\ref{eq:simpRpoles}). This suggests that the pole (\ref{eq:hydropole}) of $G_J^{0,0}(\omega,k)$ \emph{is} the hydrodynamic pole, and it exists also at intermediate coupling and wavenumber as long as $k \tau_D <\frac{\pi}{2}$. As a consequence, one can use this existence criterion as a new measure of whether a weak-coupling system behaves according to hydrodynamics at a given wave-number $k$. Furthermore, the presence or absence of the hydrodynamic pole at fixed $k$ could be used to define two different phases: an 'H phase' ('H' for hydrodynamic) for which the hydrodynamic pole is present, and a 'G phase' ('G' for gas) for which the hydrodynamic pole is absent. These two phases can be characterized by the qualitatively different real-time response $\delta n(t,k)$ at fixed wave-number $k$, where  
\begin{equation}
\delta n(t,k)\propto \int d\omega\, G_J^{0,0}(\omega,k) e^{-i \omega t}\,.
\end{equation}
Representative plots of the real-time response in the H phase ($k \tau_D<\frac{\pi}{2}$) and the G phase ($k \tau_D>\frac{\pi}{2}$) are shown in Fig.~\ref{fig:two}. As can be seen from these plots, there is a qualitative difference between the real time evolution of the density at fixed $k$ in the H and G phases, respectively. Note that, despite this qualitative change between the H and G phase, respectively, the behavior of the diffusion transport coefficient from the Kubo formula (\ref{eq:Dkubo}) is insensitive to this transition because correlation functions are evaluated in the limit of vanishing wave-number, for which the hydrodynamic pole is present at any value of the coupling. Also, the above transition between the H and G phase is a type of 'onset' transition. Somewhat more surprising is that the presence of the onset transition is extremely hard to recognize from studying the shape of the spectral function ${\rm Im}\,G_J^{0,0}(\omega,k)$ in Fourier space. 

Considering now the non-hydrodynamic signatures one realizes from the strong-coupling result Eq.~(\ref{eq:Gnnstrong}) that the branch cut does not survive the strong coupling limiting procedure. Instead, Eq.~(\ref{eq:Gnnstrong}) contains (besides the hydrodynamic pole) a number of non-hydrodynamic poles depending on the order $m$ of the approximating function $R_{m}(\omega,{\bf k})$. As the approximation order $m$ is increased, the number of non-hydrodynamic poles increase as well, and they cluster along a line in the complex $\omega$ plane from $\omega_-=\frac{i}{\tau_D}-k$ to $\omega_+=-\frac{i}{\tau_D}+k$. Specifically, one finds for small $k$:
\begin{eqnarray}
\label{eq:poles}
m=1:&\quad& \omega_1=-\frac{i}{\tau_D}\,,\nonumber\\
m=2:&\quad& \omega_{1,2}\simeq -\frac{i}{\tau_D}\pm 0.516398 k\,,\nonumber\\
m=3:&\quad& \omega_{1,2}\simeq -\frac{i}{\tau_D}\pm 0.723747 k\,,\omega_3=-\frac{i}{\tau_D}\,,\nonumber\\
m=4:&\quad& \omega_{1,2}\simeq -\frac{i}{\tau_D}\pm 0.82333 k\,,\omega_{3,4}\simeq -\frac{i}{\tau_D}\pm 0.31608 k\,,\\
\ldots \nonumber\\
m=10:&\quad& \omega_{1,2}\simeq -\frac{i}{\tau_D}\pm 0.964872 k\,,
\omega_{3,4}\simeq -\frac{i}{\tau_D}\pm0.848484k\,,\nonumber\\
&&\omega_{5,6}\simeq -\frac{i}{\tau_D}\pm 0.661558 k\,,\omega_{7,8}\simeq -\frac{i}{\tau_D}\pm 0.420034 k\,,
\omega_{9,10}\simeq -\frac{i}{\tau_D}\pm 0.143954 k\,,\nonumber
\end{eqnarray}

Not surprisingly, the clustering of these poles corresponds to the branch cut of the original function in Eq.~(\ref{eq:Gnnres}). Thus, one recovers the well-known result that a branch cut can be well approximated by a series of poles (see Fig.~\ref{fig:one}).

The R-current correlator of ${\cal N}=4$ SYM can be calculated at strong coupling using gauge/gravity calculation \cite{Kovtun:2005ev}. One does not observe a series of poles clustered along a line parallel to the real $\omega$ axis, but rather a series of pairs of poles located increasingly farther away from the real axis, of the form $\omega_{n,\pm}\propto -2\pi T(i\pm 1)n$. The poles for $n=2,3,\ldots$ can be understood to correspond to higher resonances in the quantum field theory. Specifically, these higher resonances have been shown to exist in previous calculations in quantum field theory (see e.g. Ref.~\cite{Hartnoll:2005ju}). On the other hand, kinetic theory does not include these higher resonances, because of the approximations made in deriving the kinetic equations from quantum field theory \cite{Blaizot:2001nr}. Thus, it is clear why there are no poles/cuts located at e.g. $\omega\propto -i n/\tau_D$ for $n=2,3,\ldots$ found in Eq.~(\ref{eq:Gnnres}).

However, for $n=1$, at leading order in the strong coupling calculation of ${\cal N}=4$ SYM only two poles are found for $k\neq 0$. In the corresponding 'strong coupling limit' of the kinetic theory result (\ref{eq:poles}), only one pole is found at leading order ($m=1$). It is not clear to me what is causing this qualitative discrepancy between kinetic theory and gauge/gravity duality calculations. It is possible that the discrepancy arises from the fact that the naive strong coupling limit $\tau T\rightarrow 0$ does not coincide with the actual strong coupling limit performed in quantum field theory, or because of the first-order gradient expansion in the Wigner-transform used to derive the kinetic theory from quantum field theory \cite{Blaizot:2001nr}. However, it is remarkable that when setting $\tau_D=\tau_\Delta$ and using $\tau_\Delta=\frac{\ln 2}{2\pi T}$ from gauge/gravity duality \cite{PRprep} one finds that (\ref{eq:poles}) for $n=1$ is located in the vicinity of $\omega_{1,\pm}=-2\pi T(i\pm 1)$ in strongly coupled ${\cal N}=4$ SYM.

Beyond leading order in the strong coupling expansion, the kinetic theory result for $m=2$ contains two non-hydrodynamic poles for $k\neq 0$, while the gauge-gravity calculation still contains the original two poles \cite{Waeber:2015oka}. A possible reason for this difference could be that the gauge-gravity calculation is accurate up to $\lambda^{-3/2}$ corrections, while the kinetic theory result predicts the additional pole to appear when corrections of order ${\cal O}\left(\tau_D\right)\simeq {\cal O}\left(\lambda^{-2}\right)$ are included. While there is no fundamental reason that the power counting from kinetic theory should be accurate in the strong coupling regime, it is nevertheless conceivable that a gauge-gravity calculation for the R-current correlator accurate to order $\lambda^{-3}$ would find additional poles reminiscent of the structure in Eq.~(\ref{eq:poles}). Such a finding would indeed help explain how branch cuts at weak coupling turn into poles at strong coupling in thermal correlators of large $N$ gauge theories. Alternatively, it is possible that in ${\cal N}=4$ SYM quantum field theory there is a type of onset transition at a fixed value of $k \tau_D$ where the branch cut actually dissolves into a series of poles, similar to the H-G transition found for the hydrodynamic pole above, and that the kinetic theory model is too simplistic to capture this phenomenon for the non-hydrodynamic modes. Future studies are needed to tell which scenario is realized.

\subsection{Momentum Transport}

Most features for momentum transport mirror those found for the R-charge diffusion correlator. The presence of the hydrodynamic poles, however, is somewhat different. One finds that for the spin 2 excitations $G_T^{12,12}(\omega,k)$, there simply is no pole in the correlator (\ref{eq:allhydrores}), matching previous results finding no hydrodynamic pole in this channel. For spin 1 excitations $G_T^{01,01}(\omega,k)$ there is a pole located at $\omega_0^{(1)}(k)=\frac{i(-1+y(k))}{\tau_R}$ with $y(k)$ a solution to the equation
\begin{equation}
\frac{2 k \tau_R y}{y^2-k^2 \tau_R^2}=\tan\frac{4 (k \tau_R)^3+6 k \tau_R y}{3 (k^2 \tau_R^2+y^2)}\,,
\end{equation}
subject to the existence condition $k \tau_R<\frac{3 \pi}{4}$. It is easy to verify that for $k \tau_R \ll 1$, this pole becomes the hydrodynamic pole $\omega_0^{(1)}=-\frac{i \tau_R k^2}{5}$ discussed in Eq.~(\ref{eq:spin1poles}). Hence it is natural to expect a H-G onset transition for the shear channel to occur at $k \tau_R =\frac{ 3\pi}{4}$.

For the sound (spin 0) channel correlator $G_T^{00,00}(\omega,k)$ there are two poles located at $\omega_\pm^{(0)}(k)=\frac{i(-1+y(k))\pm x(k)}{\tau_R}$ with $x(k),y(k)$ real. For small $k \tau_R$, $\omega_\pm$ once again become the usual hydrodynamic poles $\omega_{\pm}^{(0)}$ from Eq.~(\ref{eq:soundhydropoles}). The poles exist for $k \tau_R<4.5313912\ldots$, where the critical value for $k\tau_R$ is a numerical solution to the equation
\begin{eqnarray}
0&=&12 (k\tau_R)+4 (k\tau_R)^3-2 (k\tau_R)^2 \pi-6 \pi x^2+6 x \ln\left[\frac{k \tau_R-x}{k \tau_R+x}\right]\,,\\
x&=&\sqrt{\frac{-3\pi-2 (k\tau_R)^2\pi+2 k^3\tau_R^3+\sqrt{4 (k\tau_R)^6+12 (k\tau_R)^3 \pi+9 \pi^2+12 (k\tau_R)^2 \pi^2}}{6\pi}}\,.\nonumber
\end{eqnarray}
Note that the boundary between H and G phase for the spin 2 (sound) channel is at a different location than for the spin 1 (shear) channel.

\section{Summary and Conclusions}

In the present work, real time correlators for the R-charge diffusion and momentum transport have been calculated within the framework of the Boltzmann equation in the relaxation-time approximation. The results should be applicable for weakly coupled large $N$ gauge theories, but have several features that could make them useful as a model for strongly coupled gauge theories. For instance, a naive strong coupling limit, performed through sending the relaxation time to zero, gives rise to the known hydrodynamic results for the correlation functions. This is a generalization of the well-known fact that kinetic theory gives rise to the equations of classical hydrodynamics for the one-point function. In the naive strong coupling limit, the ubiquitous logarithmic branch cut found in the correlators dissolves into a series of poles, which are qualitatively similar to the non-hydrodynamic quasi-normal modes found in black holes. The naive strong coupling limit of kinetic theory also would predict that new poles appear to fill the cut order by order in powers of $\lambda^{-2}$ in the t'Hooft coupling $\lambda$, rather than the typical expansion parameter $\lambda^{-3/2}$ encountered in gauge/gravity duality.

As the coupling $\lambda$ is decreased from infinity (or equivalently if the wave-number $k$ of a perturbation is increased), the results found in this work predict a vanishing of the hydrodynamic pole(s), such that for very weak coupling only the logarithmic branch cut remains. This transition can be clearly observed through the qualitatively different real-time behavior of perturbations, and for this reason was called a 'onset transition'. It is conceivable that the concept of onset transitions could be useful to classify the transport properties in real materials. It should be possible to verify the existence of onset transitions by calculating the scanning the corresponding correlation functions in perturbative quantum field theory for arbitrary wave-number $k$. This is left for future work.

\begin{acknowledgments}

I would like to thank S.~Hartnoll, P.~Kovtun, S.~Minwalla, A.~Starinets and A.~Vuorinen for fruitful discussions.
This work was supported, in part, by the Department of Energy, DOE award No. DE-SC0008132. I also thank the Galileo Galilei Institute for Theoretical Physics for the hospitality and the INFN for partial support during the initial stages of this work.

\end{acknowledgments}

\bibliography{modes}

\end{document}